\documentclass[11pt,twoside,a4paper,cmspaper,final,collab]{cms-tdr}
\def\svnVersion{b14b060-D}\def\svnDate{2026/02/27}\def\cmsCernNoTag{CERN-EP-2026-022}\def\cmsCernDate{\today}\def\cmsMessage{Submitted to Physical Review Letters}
\begin{document}\cmsNoteHeader{SMP-25-006}

\ifthenelse{\boolean{cms@external}}{
\newcommand{\suppmat}{the supplemental material~\cite{suppMatBib}}
}{
\newcommand{\suppmat}{\appendixname~\ref{endmatter}\nocite{suppMatBib}}
}

\ifthenelse{\boolean{cms@external}}{\providecommand{\cmsLeft}{upper\xspace}}{\providecommand{\cmsLeft}{left\xspace}}
\ifthenelse{\boolean{cms@external}}{\providecommand{\cmsRight}{lower\xspace}}{\providecommand{\cmsRight}{right\xspace}}
\newcommand{\pp}{\ensuremath{\Pp\Pp}\xspace}
\newcommand{\epem}{\ensuremath{\Pep{}\Pem{}}\xspace}
\newcommand{\gtogg}{\ensuremath{\cPg\to\cPg\cPg}\xspace}
\newcommand{\gtoqq}{\ensuremath{\cPg\to\qqbar}\xspace}
\newcommand{\sqq}{\ensuremath{\text{score}_{\gtoqq}}\xspace}
\newcommand{\dphi}{\ensuremath{\Delta\varphi}\xspace}
\newcommand{\HERWIGvii}{{\HERWIG\,7}\xspace}

\cmsNoteHeader{SMP-25-006}
\title{Measurement of angular correlations inside jets induced by gluon polarization in proton-proton collisions at \texorpdfstring{$\sqrt{s} = 13.6\TeV$}{sqrt(s) = 13.6 TeV}}

\date{\today}

\abstract{A study of angular correlations inside jets induced by gluon polarization is performed using proton-proton collisions at a center-of-mass energy of $\sqrt{s}=13.6\TeV$. The data correspond to an integrated luminosity of 34.7\fbinv, collected in 2022 with the CMS detector at the LHC. The details of the parton shower are investigated using jets reconstructed with the anti-\kt algorithm and subsequently declustered with the Cambridge--Aachen algorithm. A novel analysis technique is developed to identify characteristic features of the jet substructure and to select intermediate gluon splittings into quark-antiquark pairs. An observable sensitive to gluon polarization in the parton shower is measured and compared with \PYTHIAviii and \HERWIGvii model predictions, with and without angular correlations induced by the gluon spin. The results are consistent with models that incorporate gluon polarization and strongly disfavor those that neglect them.}

\hypersetup{
pdfauthor={CMS Collaboration},
pdftitle={Measurement of angular correlations inside jets induced by gluon polarization in proton-proton collisions at sqrt(s) = 13.6 TeV},
pdfsubject={CMS},
pdfkeywords={CMS, parton shower, gluon spin}}

\maketitle 

Gluons are the gauge bosons that mediate the strong interaction, theoretically described by quantum chromodynamics (QCD). Discovered by the TASSO~\cite{TASSO:1980lqw} and PLUTO~\cite{PLUTO:1980fpl} experiments at the DESY PETRA facility, their spin-1 nature has been thoroughly established in studies of the spin correlations among jets produced in $\epem$ annihilation at the CERN LEP collider~\cite{ALEPH:1997mcm,Moretti:1998zc}. Angular correlations among the produced jets originate from the two possible polarization states of the gluon, giving rise to interference patterns that manifest themselves in spin-sensitive final-state observables. However, such patterns have never been directly observed in collinear and soft parton radiation inside the jets themselves, where the physics is usually described by parton shower (PS) models and affected by nonperturbative effects. Jets produced in multi-TeV proton-proton (\pp) collisions at the CERN LHC have much higher energies than previously reached at $\epem$ colliders. This leads to a large dynamic range for intrajet parton radiation, and makes it possible to study spin-correlation effects in the soft and collinear regime. 
Most current PS Monte Carlo (MC) event generators incorporate gluon polarization following the original Collins--Knowles approach~\cite{Collins:1987cp,Knowles:1988hu}. In the \PYTHIA~\cite{Sjostrand:2014zea} and \SHERPA~\cite{Sherpa:2019gpd} models, a modulation is added to the opening angle between the production and splitting planes of the gluon to approximate such interference patterns. In a recent \HERWIG~\cite{Richardson:2018pvo} development, spin correlations are treated explicitly using a spin-density matrix formalism. Properly describing angular correlations induced by the gluon spin, including the impact of nonperturbative effects, is crucial in modern PS algorithms that go beyond leading-logarithmic accuracy~\cite{Karlberg:2021kwr,Hamilton:2021dyz,Preuss:2024vyu,Nagy:2020rmk,Forshaw:2019ver,vanBeekveld:2022zhl}. 
Moreover, jet parton-type tagging algorithms based on simulated data rely on a realistic description of the kinematics of the particles inside jets as input. Any mismodeling of the inter-particle structure, \eg, in the jets from Higgs boson decays into two gluons where significant correlation is expected, impacts the jet tagging performance. It is thus important to measure the angular correlations of particles inside jets, check their modeling in current MC generators, and provide new inputs for their improvement.

Within a jet, the parton shower develops through successive branchings, in which gluons split into two gluons or into quark-antiquark pairs ($\Pg \to \Pg\Pg$, $\Pg \to \qqbar$) and quarks radiate gluons ($\PQq \to \PQq\Pg$). In two successive parton splittings, $\PX\to1+2$ followed by $2\to 3+4$ (where $\PX$ can be a gluon or a quark), the matrix elements of the helicity states of the intermediate parton 2 interfere and give rise to a nonuniform contribution to spin-sensitive observables. One such observable is the angle \dphi, defined as~\cite{Richardson:2018pvo,Karlberg:2021kwr}
\begin{equation}
\cos\dphi = \hat{n}_{1,2} \cdot \hat{n}_{3,4} \cdot \text{sign}  [(\hat{n}_{1,2} \times \hat{n}_{3,4}) \cdot \vec{p} ],
\label{eq:dphi}
\end{equation}
where $\hat{n}_{1,2}$ and $\hat{n}_{3,4}$ are unit vectors of the planes formed by partons 1, 2 and partons 3, 4, respectively, and $\vec{p}$ is the momentum of the most energetic of partons 1 and 2. Theoretical calculations~\cite{Richardson:2018pvo} predict only spin correlations in $\gtogg$ and $\gtoqq$ branchings, leading to a characteristic \dphi modulation.
This modulation is much more pronounced in the $\gtoqq$ than in the $\gtogg$ branchings. However, since the $\gtogg$ process has a larger rate and the two modulations have opposite signs, the overall spin effect largely cancels out in the shower. Therefore, the study of gluon spin correlations requires separating the different processes and selecting the $\gtoqq$ contribution in the second branching.

This Letter reports on the measurement of the distribution of the angle \dphi between two successive parton radiation processes inside jets using \pp collision data collected by the CMS experiment in 2022 (Run~3), corresponding to an integrated luminosity of 34.7\fbinv. This is the first study of spin correlations induced by gluon polarization inside jets. Jets are reconstructed with the anti-\kt algorithm~\cite{Cacciari:2008gp,Cacciari:2011ma} with a distance parameter 0.8, denoted as AK8, and declustered using the Cambridge--Aachen (CA) algorithm~\cite{Dokshitzer:1997in}. The chains of declustered subjets are regarded as proxies for the different parton branching processes. Two subsequent declustered subjets are selected for each jet to calculate the angle \dphi according to Eq.~(\ref{eq:dphi}). A machine-learning (ML) approach is used to categorize events by parton-splitting topology and to select those most likely originating from $\PX\to \PX \Pg$ and \gtoqq branchings. 
The measured \dphi distributions are compared with MC simulations with gluon polarization included (correlation-on) and neglected (correlation-off) to evaluate the size of the effect and examine the accuracy of the PS modeling.

The CMS apparatus~\cite{CMS:2008xjf,CMS:2023gfb} is a multipurpose, nearly hermetic detector, designed to trigger~\cite{CMS:2020cmk,CMS:2016ngn,CMS:2024aqx} on and identify electrons, muons, photons, and (charged and neutral) hadrons~\cite{CMS:2020uim,CMS:2018rym,CMS:2014pgm}. A global ``particle-flow'' algorithm~\cite{CMS:2017yfk} aims to reconstruct all individual particles in an event, combining information provided by the all-silicon inner tracker and by the crystal electromagnetic and brass-scintillator hadron calorimeters, operating inside a 3.8\unit{T} superconducting solenoid, with data from the gas-ionization muon detectors embedded in the flux-return yoke outside the solenoid. The reconstructed particles are used to build \Pgt leptons, jets, and missing transverse momentum~\cite{CMS:2018jrd,CMS:2016lmd,CMS:2019ctu}.
Events of interest are selected using a two-tiered trigger system. The first level, composed of custom hardware processors, uses information from the calorimeters and muon detectors to select events at a rate of around 100\unit{kHz} within a fixed latency of 4\mus~\cite{CMS:2020cmk}. The second level, known as the high-level trigger (HLT), consists of a farm of processors running a version of the full event reconstruction software optimized for fast processing, and reduces the event rate to a few kHz before data storage~\cite{CMS:2016ngn,CMS:2024aqx}.

The jet momentum is determined as the vectorial sum of the momenta of all particles clustered into the jet and, according to simulation, reproduces the generator-level momentum on average to within 5--10\% over the full \pt range and detector acceptance. Additional \pp interactions within the same or nearby proton bunch crossings (pileup) can contribute extra tracks and calorimetric energy depositions, increasing the apparent jet momentum. The pileup per particle identification algorithm~\cite{Sirunyan:2020foa,Bertolini:2014bba} is used to mitigate this effect at the reconstructed-particle level, making use of local shape information, event pileup properties, and tracking information. The primary vertex is taken to be the vertex corresponding to the hardest scattering in the event, evaluated using tracking information alone, as described in Section 9.4.1 of Ref.~\cite{CMS-TDR-15-02}. Charged particles identified to be originating from pileup vertices are discarded. The momenta of the neutral particles are rescaled according to their probability to originate from the primary interaction vertex, deduced from the activity surrounding the particle~\cite{Sirunyan:2020foa}. 
To obtain a sample of jet-enriched events with abundant parton splittings, jets of large size and high \pt are preferred. Events are selected at the HLT by requiring the presence of a jet with $\pt >550\GeV$, corresponding to greater than 99.5\% efficiency for jets with $\pt> 700\GeV$ according to simulation. The reconstructed AK8 jets are required to have $\pt>700\GeV$ and $\abs{\eta} < 1.7$. The events must contain at least two jets separated in azimuth by two radians or more. 

Simulations of dijet events are generated at leading order (LO) with \PYTHIA{\,8.306}~\cite{Bierlich:2022pfr} with \pt-ordered shower and the CP5 underlying event tune~\cite{CMS:2019csb}, hereafter called \PYTHIAviii, as well as with \HERWIG{\,7.3.0}~\cite{Bahr:2008pv,Bellm:2015jjp} with angular-ordered shower and the CH3 tune~\cite{CMS:2020dqt}, hereafter called \HERWIGvii. Both MC generators use the NNPDF31\_NNLO parton distribution functions (PDFs)~\cite{NNPDF:2017mvq}. The full detector response is simulated with \GEANTfour~\cite{Agostinelli:2002hh}.

The CA algorithm clusters particle pairs or subjets closest in angle sequentially, mimicking the angular ordering of LO parton splittings. 
In this analysis, charged and neutral particles with $\pt>1\GeV$ inside jets are used to reconstruct two successive parton splittings. 
The clustering of particles and subjets is repeated until all particles are recombined in a jet, thereby providing a proxy for the parton splitting history. For each jet, two successive branchings ($\PX\to1+2$ and $2\to 3+4$) are selected from the chain. The method proposed in Ref.~\cite{Karlberg:2021kwr} is followed, whereby subjet 1 and 2 (3 and 4) are the harder and softer subjets of the highest \kt splitting in the primary (secondary) Lund plane~\cite{Dreyer:2018nbf}, respectively, with
the $\kt$ variable defined as
\begin{equation}
	\kt=\pt^{\text{soft}}\Delta R_{i,j},\quad\Delta R_{i,j}=\sqrt{(y_i-y_j)^2+(\phi_i-\phi_j)^2},
    \label{eq:DefitionOfkt}
\end{equation}
where $\pt^{\text{soft}}$ is the transverse momentum of the soft subjet in the splitting, $\Delta R_{i,j}$ represents the angular distance between subjets $i$ and $j$, and $y_i$ and $\phi_i$ are the rapidity and azimuthal angle of subjet $i$. An extra $\pt>130\GeV$ requirement is applied for the subjet 2 to reduce the impact of softer radiations. No additional thresholds on the \pt of subjet 3 and 4 are required.

The selected subjets in the second branching, \ie, subjets 3 and 4, may originate from quarks or gluons. To identify the type of the reconstructed subjets, they are matched in the simulation to the subjets made of final-state MC-level partons with $\pt>1 \GeV$. These parton subjects are obtained via CA declustering as well, and subjet 3 and 4 can be selected following the same method selecting reconstructed subjets. The definition of the parton subjet type can be ambiguous since the particles inside them do not necessarily originate from the same splitting, and can be affected by soft and collinear radiation. The interleaved flavor neutralisation~\cite{Caola:2023wpj} prescription is applied, which provides infrared- and collinear-safe definitions of the jet type at each stage of the clustering sequence. This is realized by a special treatment of soft \qqbar pairs. In ideal cases, the type of a subjet is either one of quarks or a gluon, but the CA declustering might also include partons from other branchings. This happens in about 1\% of the cases in the study, and results in a nonsinglet flavor, such as $\PQu\PAQd$, referred to as the ``rest'' type. 

Reconstructed and parton subjets are matched if their distance is less than half of $\Delta R_{3,4}$, the separation between the reconstructed subjets 3 and 4. Typically, $\Delta R_{3,4}$ is around 0.1. If either one of subjets 3 and 4 is not matched to a simulated parton-level subjet, the pair is denoted as ``unmatched''. With this approach, the fractions of $\PQq \Pg$, $\Pg\Pg$, \qqbar, ``unmatched'', and ``rest'' types are found to be about 7\%, 24\%, 4\%, 65\%, and 0\%, respectively, in the \HERWIGvii simulation. The \PYTHIAviii results show a similar composition of 8\%, 16\%, 3\%, 72\%, and 1\%, respectively.
The unmatched subjets have typically more collinear and softer activity, with $\Delta R_{3,4}<0.05$ and $\pt<15\GeV$ for subjet 4. Alternatively, one can obtain the type by matching subjets 2 and 3. In this case, the above fractions become 18\%, 52\%, 6\%, 23\%, and 1\% for \HERWIGvii;  and 24\%, 45\%, 6\%, 23\%, and 2\% for \PYTHIAviii simulations, respectively.
The much smaller unmatched rate than in the previous case indicates that nonperturbative effects in the collinear and soft region, as well as limited detector resolution, have a larger impact on subjet 4, and consequently smear the \dphi distribution.

Since the spin correlation effect is most pronounced in the $\PX\to \PX \Pg$ with $\gtoqq$ topologies, which account for a small fraction of all events, ML techniques are applied to improve their selection. The labeled subjets 3 and 4 in \HERWIGvii correlation-off model are used to train a deep neural network (DNN) to categorize reconstructed data in four classes: $\PQq \Pg$, $\Pg\Pg$, \qqbar, and unmatched. The fraction of the ``rest'' type of events is negligible, and therefore they are not used in the training. Variables describing the internal structure of subjets 3 and 4, as well as the splitting features of the first and second branching are used for the training. 
In total, twelve variables are included: the transverse momentum fraction of the subjet trailing in \pt in the first ($z_1$) and second ($z_2$) branching;
the separation $\Delta R_{3,4}$ between subjets 3 and 4;
the $\kt$ of the second $2 \to 3+4$ branching;
the numbers of charged and any particles in subjets 3 and 4;
the \pt-weighted width of subjets 3 and 4, $\sigma = {\sum_{i} \pt^i \Delta R^i}/{\sum_{i} \pt^i}$ 
(where $\pt^i$ is the \pt of particle $i$ inside a subjet, and $\Delta R^i$ its angular distance to 
the subjet axis in the $\eta$-$\phi$ plane);
and the \pt dispersion of subjets 3 and 4, $\pt^\mathrm{D} = \sqrt{{\sum_{i} (\pt^i) ^2}}/{\sum_{i} \pt^i}$.

\begin{figure}[htbp]
\centering
\includegraphics[width=0.48\textwidth]{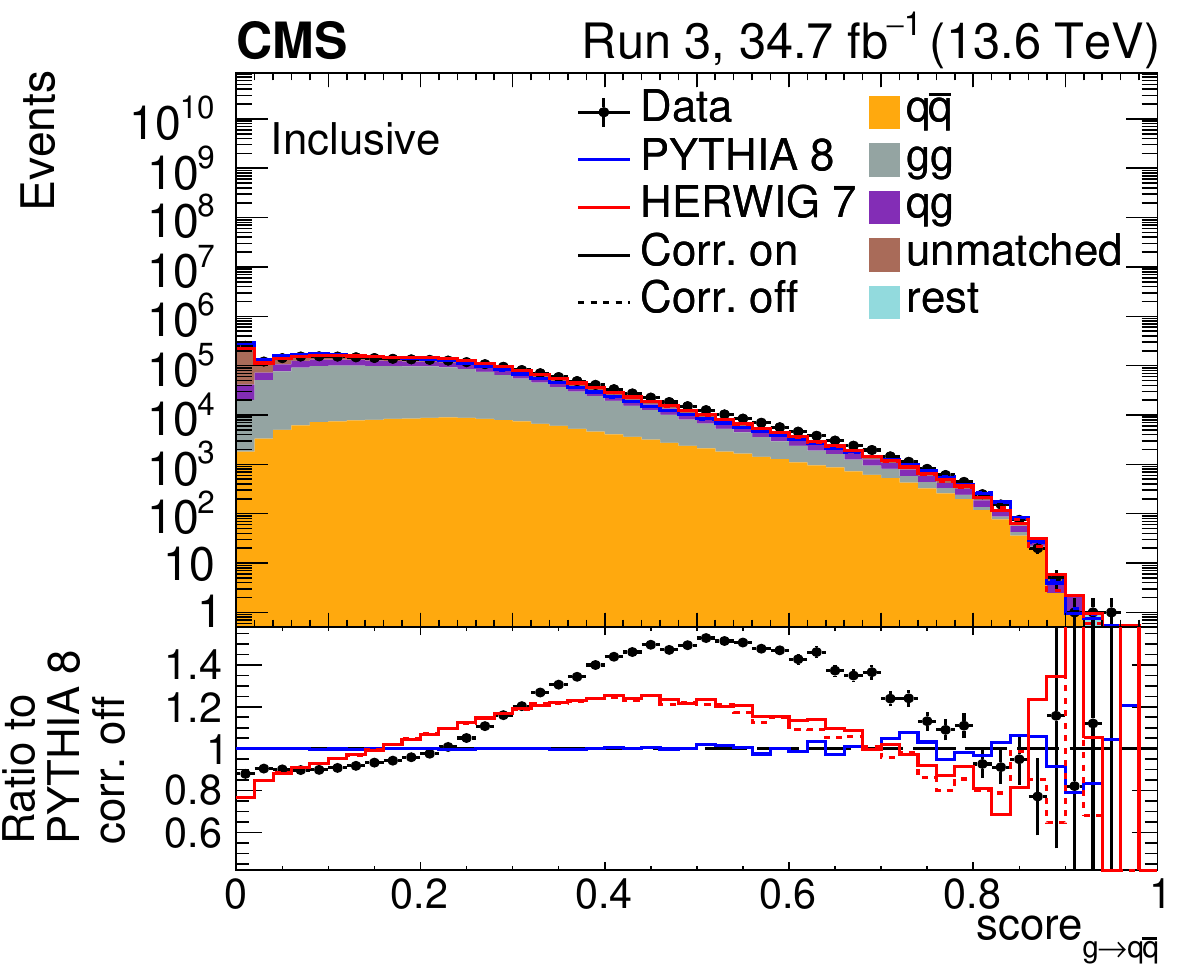}
\caption{Distribution of the DNN output score, before reweighting, for a jet identified in the \qqbar class in inclusive events (\ie, before any categorization) in data (black dots with error bars indicating statistical uncertainties) and in the different MC simulations (histograms, indicating different parton splitting contributions). Solid and dashed lines represent correlation-on and -off models, respectively. The lower panel shows the ratio of data and various MC models over the \PYTHIAviii correlation-off prediction.\label{fig:scoreNorm}}
\end{figure}

The DNN output consists of four scores, each representing the probability of a jet originates from one of the four decay final states. We denote the score for the \qqbar final state as `\sqq'. The `\sqq' distribution for the \qqbar class in data and MC simulations is shown in Fig.~\ref{fig:scoreNorm} for inclusive events (\ie, before categorization). This score distribution provides good discrimination among the different processes, with the \qqbar process dominating at high values. The correlation-on and -off models have almost identical score distributions, which ensures the independence of the DNN score from the spin correlation effects. Neither \PYTHIAviii nor \HERWIGvii simulations describe data well in the score distribution, due to the mismodeling of different input variables. A reweighting is applied to MC events to match the data. Since there are four output scores and they add up to unity for each jet, only three of them are independent. Therefore, a three-dimensional reweighting is applied hereafter to the \qqbar, $\Pg\Pg$, and unmatched scores. 
After reweighting, the MC distributions of the input variables are consistent with the data within 10\%. Events are then classified into four categories according to their highest score among them. About 50\% of events are classified into the unmatched category, with a 42\% purity of true unmatched subjets. Here, for the identification of the true type of the splitting, the matching of subjet 2 and 3 is used. The $\Pg\Pg$ category takes up about 20\% of the events with a purity of 75\%. The \qqbar category represents 12\% of the inclusive events, with a 15\% purity. 

\begin{figure}[htbp]
\centering
\includegraphics[width=0.48\textwidth]{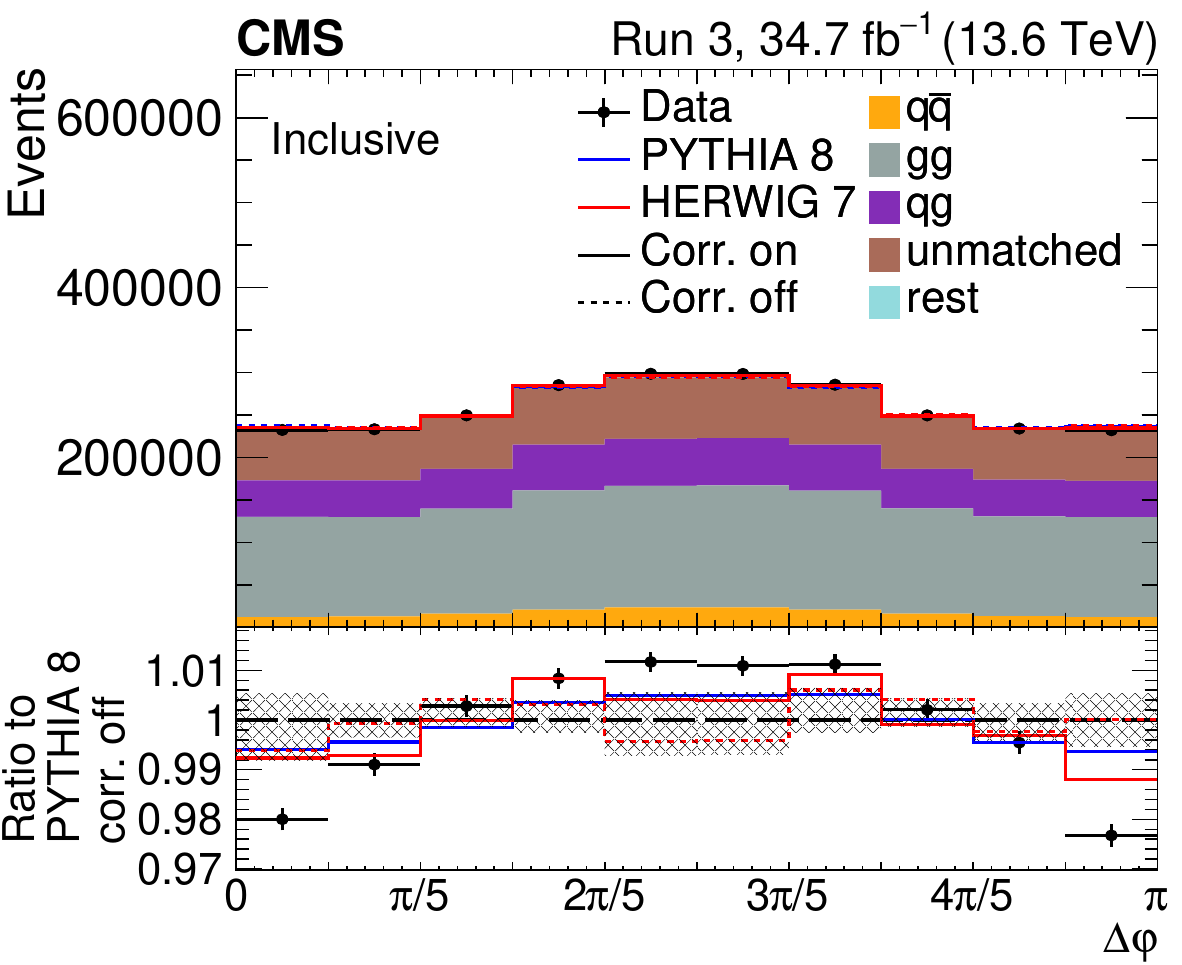}
\includegraphics[width=0.48\textwidth]{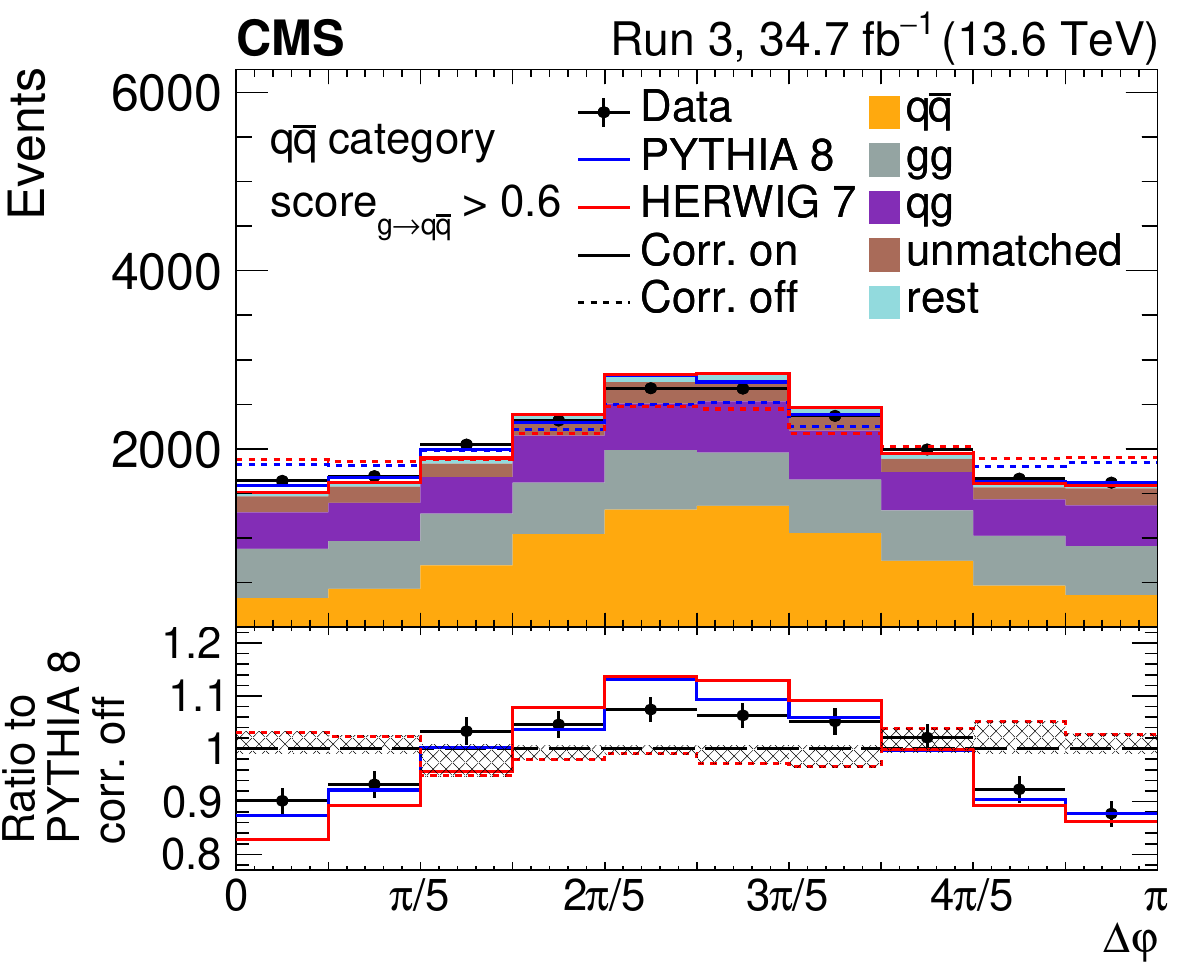}
\caption{Reconstructed \dphi distributions in data (black dots with error bars indicating statistical uncertainties) and MC simulations (histograms, indicating different parton splitting contributions) for the inclusive sample (\cmsLeft), and in the \qqbar category with $\sqq > 0.6$ (\cmsRight). The $\PQq,\Pg$ composition breakdown is shown for the \HERWIG simulation. The lower panels show the ratio of the data to the \PYTHIAviii correlation-off prediction. Systematic uncertainties are shown as gray hatched bands.\label{fig:phi_datamcCate}}
\end{figure}

Figure~\ref{fig:phi_datamcCate} (\cmsLeft) shows the reconstructed \dphi distributions of inclusive events passing all selections compared with various MC predictions. The difference between the correlation-on and -off models is very small, about 0.5\%, due to the cancellation of polarization effects in the $\gtogg$ and $\gtoqq$ processes and to the presence of a large portion of other event types without spin correlations. The \PYTHIAviii and \HERWIGvii models predict similar shapes. The data agree better with the correlation-on model, with a slightly larger modulation. The \dphi distributions in the unmatched, $\PQq\Pg$, and $\Pg\Pg$ categories can be found in \suppmat. They feature small spin correlations and serve as flavor-enriched regions for the comparison with MC simulations. The events in the \qqbar category are used to study the impact of the gluon polarization effects. With increasing \qqbar score requirement, the spin correlation effect is enhanced at the cost of event losses, as shown in \suppmat. An optimization as a function of the \qqbar score selection is performed to maximize the significance against the correlation-off hypothesis. The test statistic $q$ is defined as twice the negative log-likelihood ratio of the correlation-off and correlation-on hypotheses. Events with $\sqq>0.6$ are thus selected, which enhances the purity of the \qqbar class to 38\%, as well as the impact of gluon polarization effects. The differences between predictions with and without spin correlations are significantly larger in the \qqbar category (Fig.~\ref{fig:phi_datamcCate}, \cmsRight) than in the inclusive one (Fig.~\ref{fig:phi_datamcCate}, \cmsLeft). A maximal 12\% difference is seen in the \dphi distribution between the correlation-on and -off \PYTHIAviii predictions, whereas \HERWIGvii shows stronger gluon-spin-induced correlations. Data favor the correlation-on models with a slightly smaller modulation than predicted. 

Uncertainties arising from missing higher-order corrections in the hard scattering and parton shower are estimated with a seven-point method, from the envelope obtained by independently varying renormalization and factorization scales ($\mu_\mathrm{R}$ and $\mu_\mathrm{F}$) by factors of 2, 1, and 0.5 times the relevant energy scale, with the ($\mu_\mathrm{R}=2$, $\mu_\mathrm{F}=0.5$) and ($\mu_\mathrm{R}=0.5$, $\mu_\mathrm{F}=2$) combinations removed from the total nine combinations. The uncertainty associated with the proton PDFs is evaluated using the envelope of variations obtained with the 100 NNPDF replica sets~\cite{NNPDF:2017mvq}. The MC generators differ in their PS and hadronization models, as well as in the treatment of angular correlations induced by the gluon spin, with \PYTHIAviii using an average effect and \HERWIGvii tracking explicitly the helicity evolution in the shower. The \PYTHIAviii predictions are used as a baseline, and their differences with respect to \HERWIGvii are treated as a single-sided shape uncertainty. This model difference together with the $\mu_\mathrm{R,F}$ scale variations are the largest sources of theoretical systematic uncertainties, which amount to 0.5--3\% and 0.5--1\% shape effects in the \dphi distributions, respectively. The PDF uncertainty is smaller than 0.1\%.

Experimental uncertainties include the jet energy scale and resolution, as well as the energy scale of the jet constituents. The latter are 3\% for photons, 5\% for neutral hadrons, and 1\% for charged particles~\cite{CMS:2024mlf}. The event and subjet selections are repeated after varying the \pt of the jet and its constituents within their uncertainties. The track reconstruction efficiency uncertainty is 3\%, due to mismodeling of charged-particle tracks in the dense core of the jets~\cite{CMS:2016lmd}. The pileup uncertainty is obtained by varying the inelastic \pp\ cross section used to generate pileup interactions in the simulation.
The uncertainty in the DNN score reweighting is assessed by varying the \PYTHIAviii MC flavor composition to that of \HERWIGvii, via subjet type-dependent scale factors applied to each event. The finite size of the simulated event samples introduces statistical fluctuations in the \dphi distribution, and Poisson uncertainties are added per bin, at the scale of 1\%, to account for it. The largest experimental systematic uncertainties are the neutral particle energy scale and pileup, both have a 0.2--1\% shape effect on the \dphi distribution. 

The \textsc{Combine} tool~\cite{CMS:2024onh} is used to obtain a statistical interpretation of the observed angular correlations induced by the gluon spin in data. The two hypotheses, correlation-on and -off, are tested against data using \dphi distributions of the events with $\sqq>0.6$. Pseudodata are generated for the two hypotheses' shapes with the \PYTHIAviii simulation including systematic variations. The observed (expected) significances against the correlation-off and correlation-on hypotheses are $6.8~(8.1)$ and $1.9$ standard deviations, respectively. 

\begin{figure}[htbp]
\centering
\includegraphics[width=0.48\textwidth]{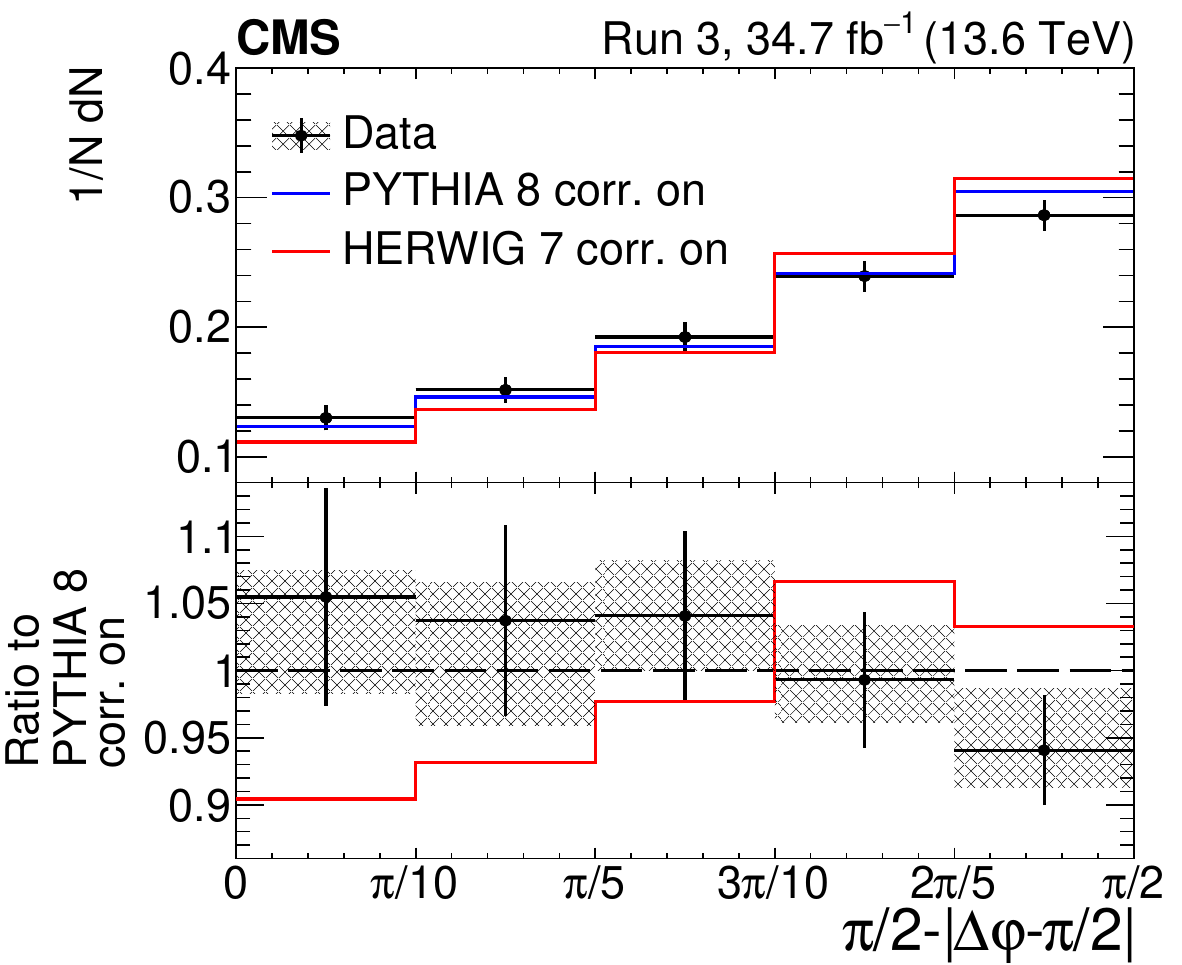}
\caption{Unfolded \dphi distribution in events of the \qqbar category with $\sqq > 0.6$ in data (black dots with error bars indicating
statistical uncertainties) compared with \PYTHIAviii and \HERWIGvii correlation-on predictions (histograms) passing the event selection described in the text. The lower panel shows the ratio of the data and the \HERWIG (correlation-on) model over the \PYTHIAviii correlation-on prediction. Systematic uncertainties are shown as gray hatched bands.\label{fig:unfolded}}
\end{figure}

The reconstructed \dphi distribution is unfolded to correct for instrumental effects. Exploiting its symmetry around $\pi/2$, the symmetrized $\pi/2 - \abs{\dphi - \pi/2}$ distribution is unfolded to reduce statistical uncertainties. The particle-level phase space definition requires at least two AK8 jets with $\pt>600\GeV$ and $\abs{\eta}<1.7$. All particles with $\pt>1 \GeV$ are used for CA declustering, and the highest-$\kt$ splittings in the primary and secondary Lund planes are selected. Subjet 2 must have $\pt>120\GeV$ and the detector-level $\Pg \to \qqbar$ DNN is applied to particle-level subjets with $\sqq>0.6$. Events passing both particle- and detector-level selections (with about $40\%$ overlap) define the migration matrix, with bin-to-bin migrations below 10\%. 
Uncertainties described above are propagated through the unfolding and added in quadrature in each bin. 
An extra modeling uncertainty, obtained from the difference between the unfolded \PYTHIAviii and \HERWIGvii results, is added. 
The leading systematic uncertainties are due to the unfolding modeling (1--7\%) and tracking efficiency (1--3\%). 
The unfolded distribution is shown in Fig.~\ref{fig:unfolded}. Statistical and systematic uncertainties both range from 4\% to 8\%. 
The data agree within uncertainties with the \PYTHIAviii correlation-on prediction, and show larger deviations with respect to the \HERWIGvii correlation-on model, as already seen in the detector-level distributions. Tabulated results, together with a \textsc{Rivet} routine to extract the DNN score, are provided in the HEPData record for this analysis~\cite{hepdata}.

In summary, this Letter reports on the first observation of angular correlations induced by the gluon polarization in soft and collinear parton emissions inside jets. The analysis uses proton-proton collision data at a center-of-mass energy of $\sqrt{s}=13.6\TeV$, 
corresponding to an integrated luminosity of 34.7\fbinv, collected in 2022 with the CMS detector at the LHC. 
The details of the parton shower are studied using jets reconstructed with the anti-\kt algorithm and declustered with the 
Cambridge--Aachen algorithm. A novel technique has been developed to identify characteristic features of the jet substructure and 
to select intermediate gluon splittings into quark-antiquark pairs. The angle between the gluon's production and decay planes is measured and compared with \PYTHIAviii and \HERWIGvii predictions, with and without spin correlations. The data deviate from the uncorrelated and spin-correlated hypotheses by 6.8 and 1.9 standard deviations, strongly disfavoring predictions that do not account for the gluon polarization. The spin correlations are apparent even in the presence of nonperturbative final-state effects, highlighting the importance of including the impact of gluon polarization in the event generators. The current models do not fully reproduce the shape of the angular modulation observed in the distributions sensitive to the gluon spin, and the unfolded measurement allows further improvements in the parton shower models. The study also presents the first identification of light-quark and gluon subjets from intermediate parton 
emissions within jets in data, using collinear- and infrared-safe flavor definitions. The proposed tagging method can be extended to the full jet splitting evolution, enabling future studies of flavor-dependent parton dynamics in the Lund plane and of spin correlations in gluon splittings, such as those relevant to Higgs boson decays into two gluons.

\begin{acknowledgments}
We congratulate our colleagues in the CERN accelerator departments for the excellent performance of the LHC and thank the technical and administrative staffs at CERN and at other CMS institutes for their contributions to the success of the CMS effort. In addition, we gratefully acknowledge the computing centers and personnel of the Worldwide LHC Computing Grid and other centers for delivering so effectively the computing infrastructure essential to our analyses. Finally, we acknowledge the enduring support for the construction and operation of the LHC, the CMS detector, and the supporting computing infrastructure provided by the following funding agencies: SC (Armenia), BMBWF and FWF (Austria); FNRS and FWO (Belgium); CNPq, CAPES, FAPERJ, FAPERGS, and FAPESP (Brazil); MES and BNSF (Bulgaria); CERN; CAS, MoST, and NSFC (China); MINCIENCIAS (Colombia); MSES and CSF (Croatia); RIF (Cyprus); SENESCYT (Ecuador); ERC PRG and PSG, TARISTU24-TK10 and MoER TK202 (Estonia); Academy of Finland, MEC, and HIP (Finland); CEA and CNRS/IN2P3 (France); SRNSF (Georgia); BMFTR, DFG, and HGF (Germany); GSRI (Greece); NKFIH (Hungary); DAE and DST (India); IPM (Iran); SFI (Ireland); INFN (Italy); MSIT and NRF (Republic of Korea); MES (Latvia); LMTLT (Lithuania); MOE and UM (Malaysia); BUAP, CINVESTAV, CONACYT, LNS, SEP, and UASLP-FAI (Mexico); MOS (Montenegro); MBIE (New Zealand); PAEC (Pakistan); MES, NSC, and NAWA (Poland); FCT (Portugal);  MESTD (Serbia); MICIU/AEI and PCTI (Spain); MOSTR (Sri Lanka); Swiss Funding Agencies (Switzerland); MST (Taipei); MHESI (Thailand); TUBITAK and TENMAK (T\"{u}rkiye); NASU (Ukraine); STFC (United Kingdom); DOE and NSF (USA).
\end{acknowledgments}\section*{Data availability} Release and preservation of data used by the CMS Collaboration as the basis for publications is guided by the  \href{https://doi.org/10.7483/OPENDATA.CMS.1BNU.8V1W}{CMS data preservation, re-use and open access policy}.

\bibliography{auto_generated} 

\ifthenelse{\boolean{cms@external}}{}{
\clearpage
\appendix
\numberwithin{figure}{section}
\section{Additional analysis figures}\label{endmatter}

Figure~\ref{fig:phiOther} shows the \dphi distribution reconstructed in data and in the MC simulations for the unmatched (upper left), $\PQq \Pg$ (upper right), and $\Pg\Pg$ (lower) categories.
\begin{figure*}[htbp]
\centering
\includegraphics[width=0.45\linewidth]{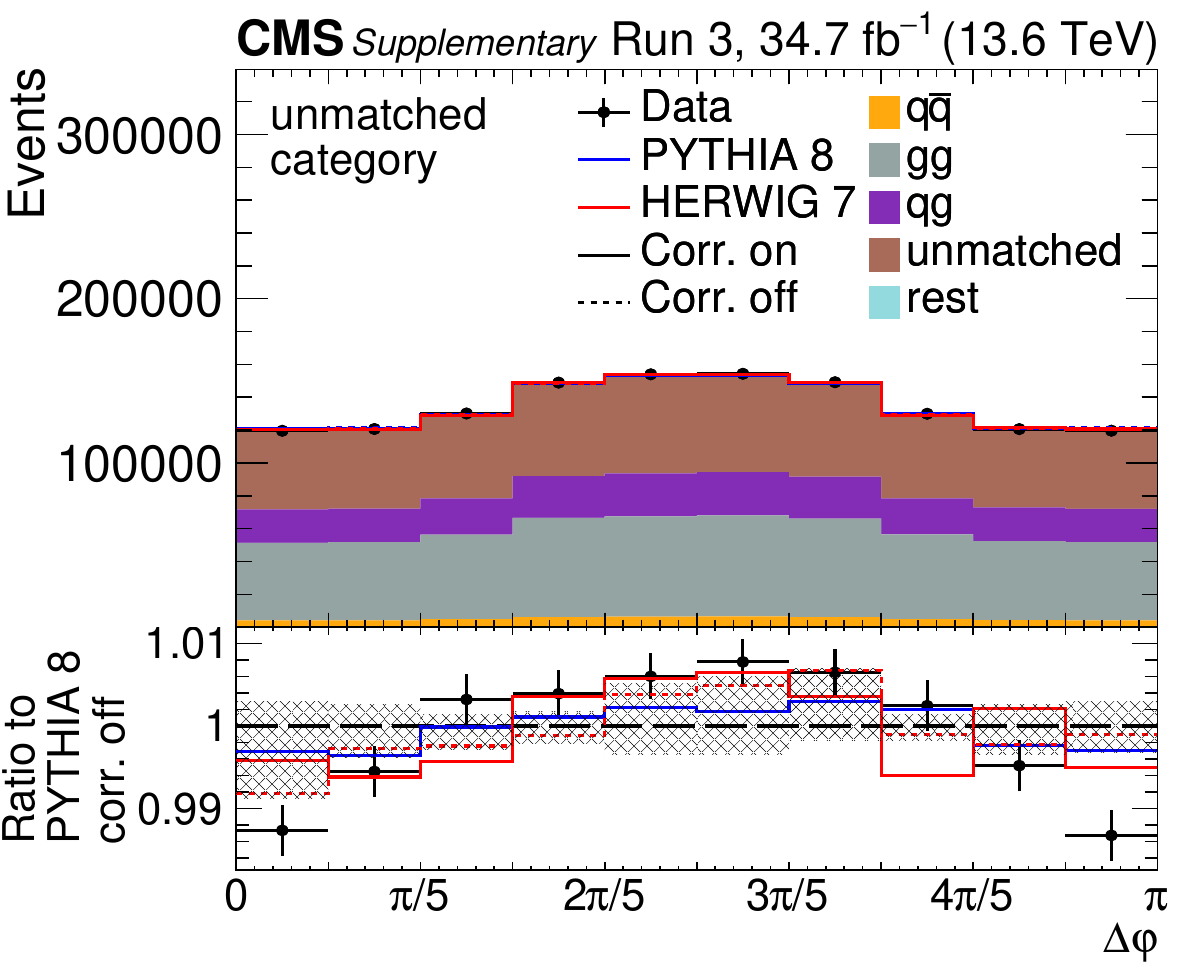}
\includegraphics[width=0.45\linewidth]{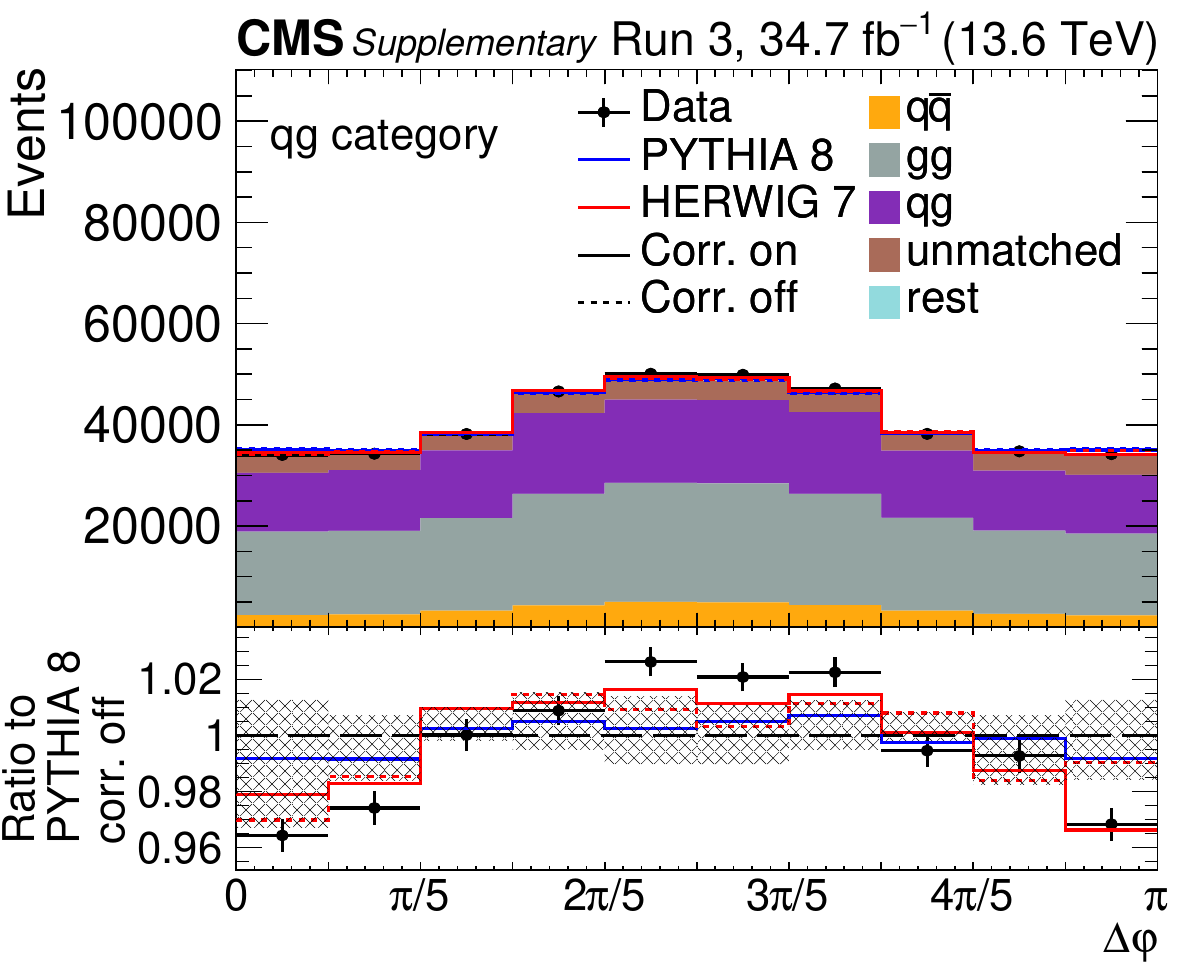}
\includegraphics[width=0.45\linewidth]{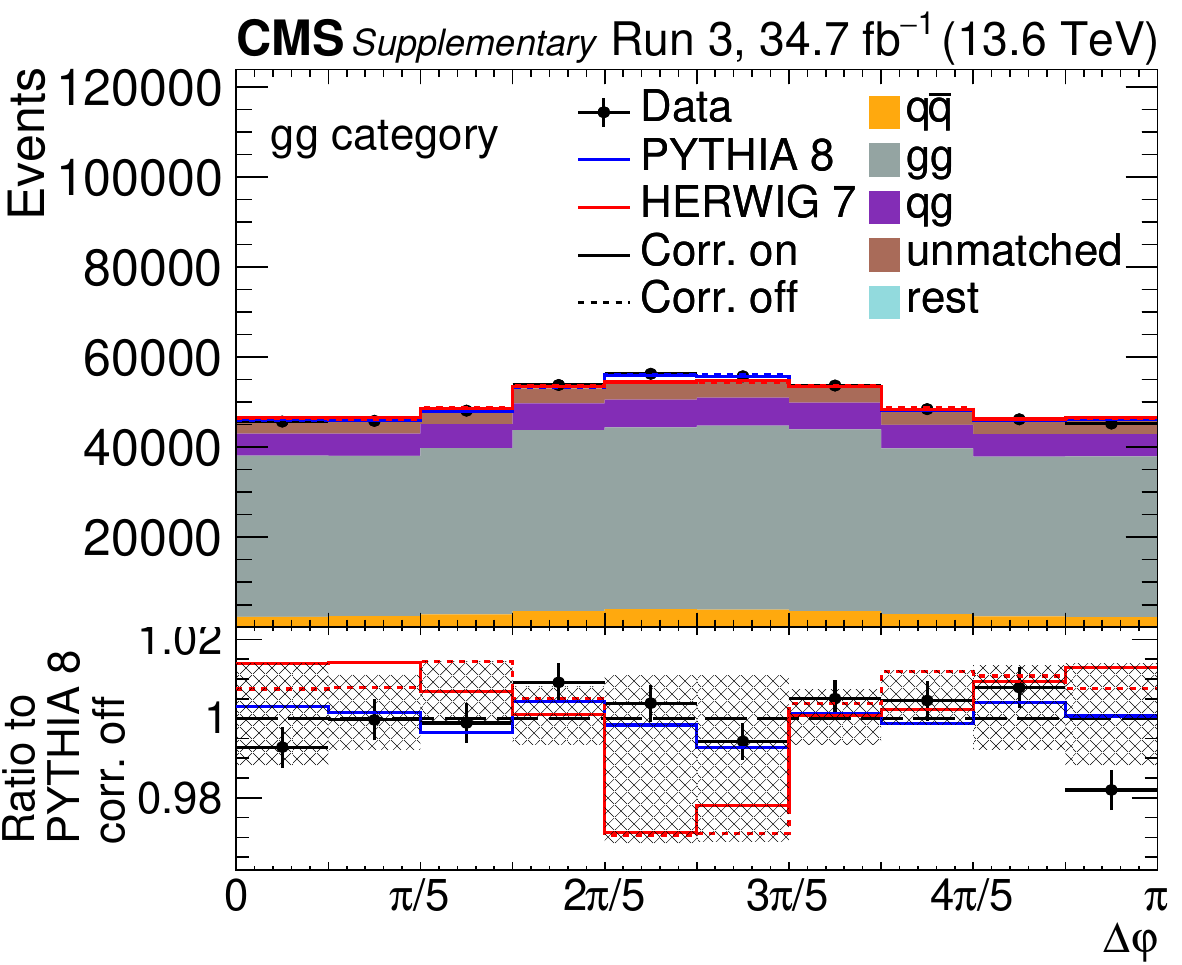}
\caption{Reconstructed \dphi distributions in data (black dots with error bars indicating statistical uncertainties) compared with the \PYTHIAviii and \HERWIGvii predictions (histograms) in the unmatched (upper left), $\PQq \Pg$ (upper right), and $\Pg\Pg$ (lower) categories.  The lower panels show the ratio of the data to the \PYTHIAviii correlation-off prediction. Systematic uncertainties are shown as gray hatched bands.\label{fig:phiOther}}
\end{figure*}

Figure~\ref{fig:phiQQ} shows the reconstructed \dphi distribution in data compared with MC simulations in the \qqbar category with two different score selections. In addition, the lower panel shows unfolded \dphi distributions for the \qqbar category for various score selections measured in data. 

\begin{figure*}[htbp]
\centering
\includegraphics[width=0.45\linewidth]{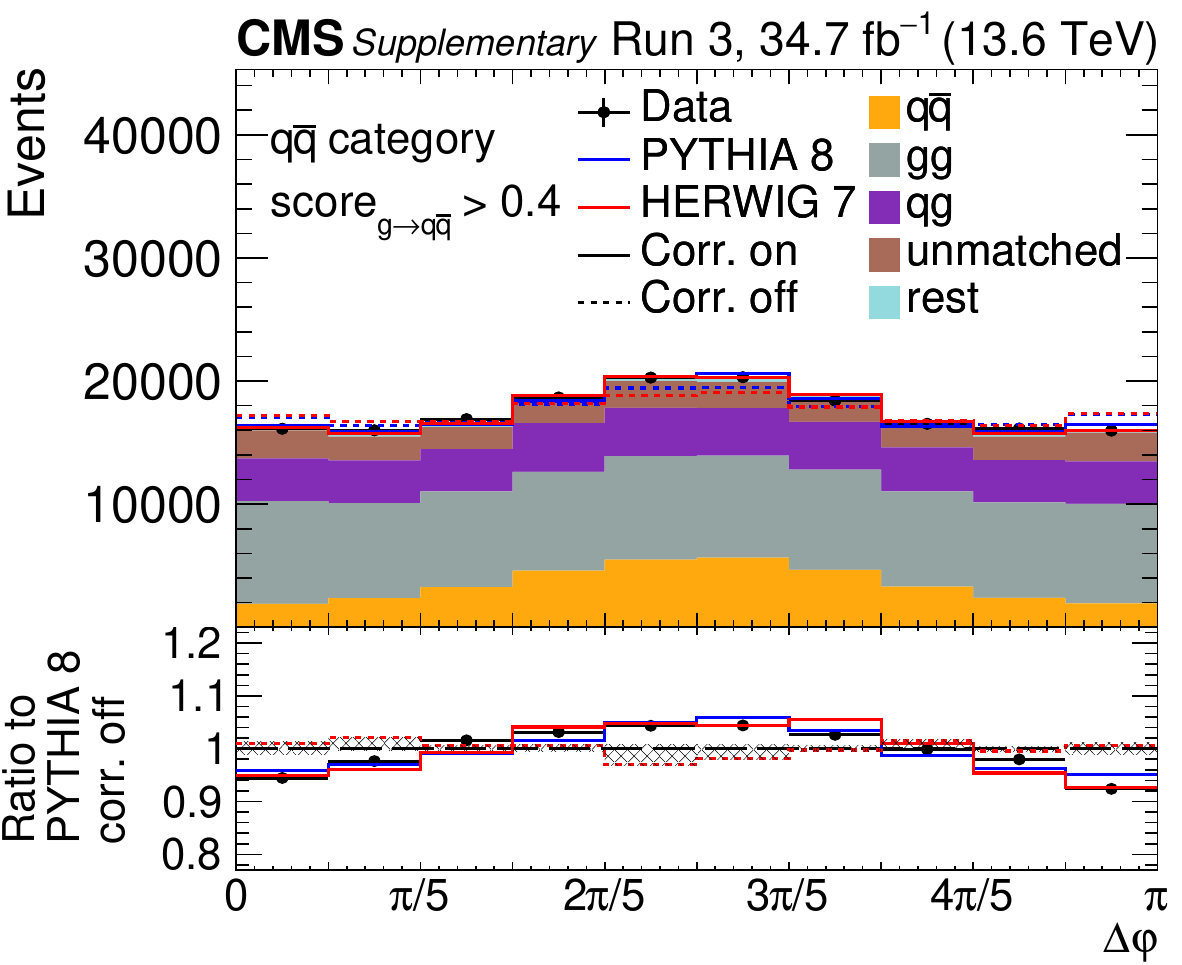}
\includegraphics[width=0.45\linewidth]{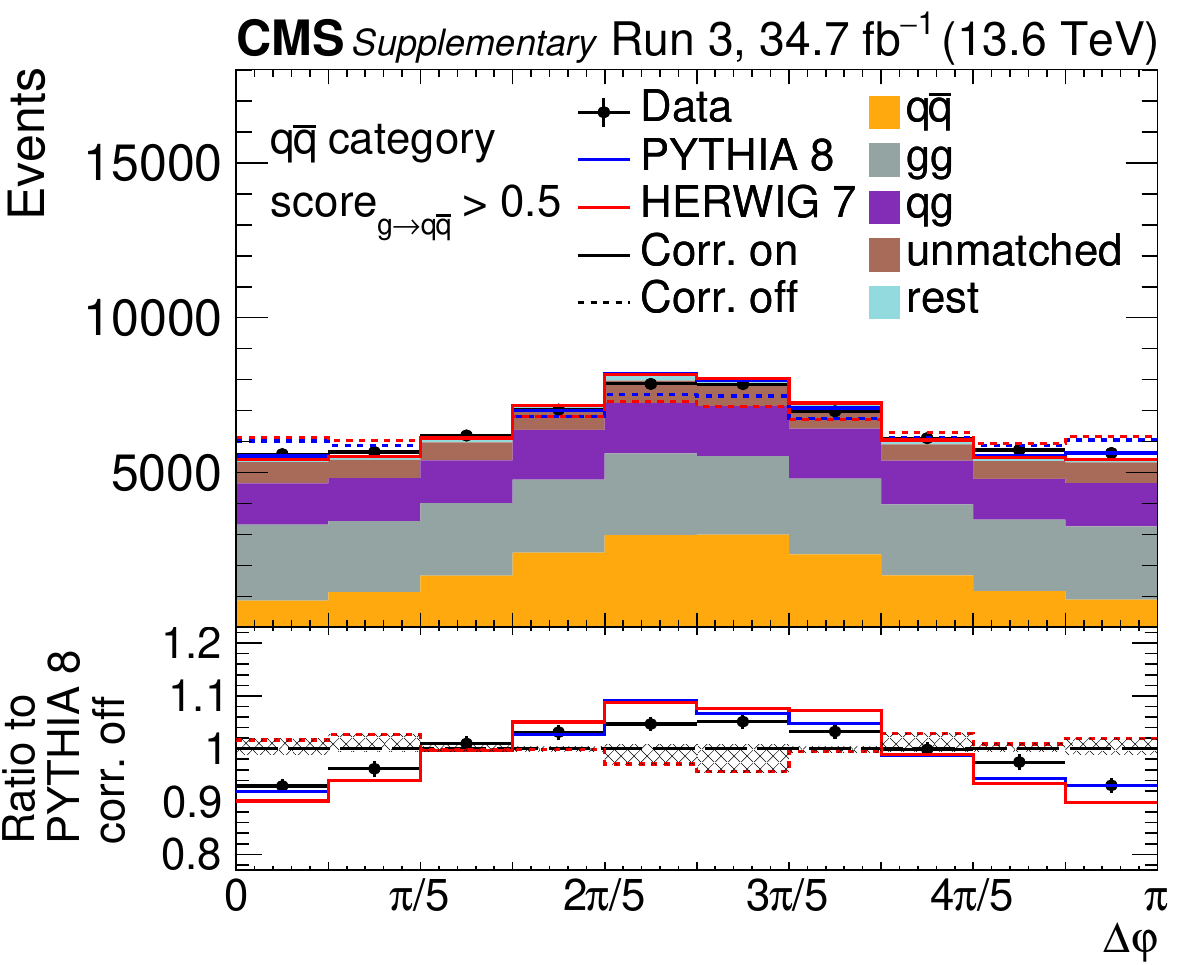}
\includegraphics[width=0.45\linewidth]{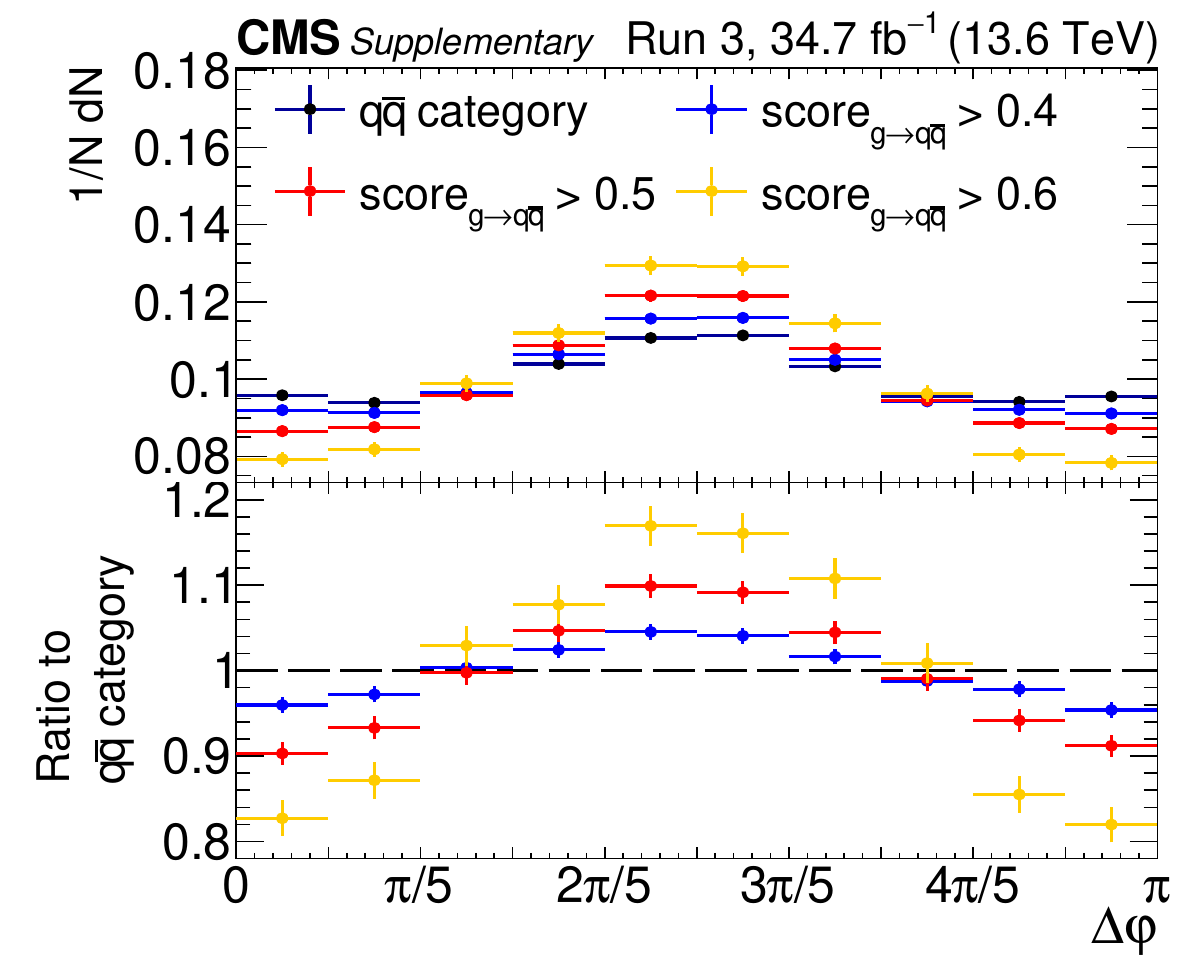}
\caption{Reconstructed \dphi distributions in data (black dots with error bars indicating statistical uncertainties) compared with the \PYTHIAviii and \HERWIGvii predictions (histograms) in the \qqbar category with scores above 0.4 (upper left) and 0.5 (upper right) selections. The lower panels show the ratio of the data to the \PYTHIAviii correlation-off prediction. Systematic uncertainties are shown as gray hatched bands. The lower plot shows unfolded \dphi distributions for the \qqbar category for various score selections measured in data.\label{fig:phiQQ}}
\end{figure*}

Figure~\ref{fig:unfold_inclusive} shows the unfolded \dphi distribution in the phase space without any score selections. 

\begin{figure*}[htbp]
\centering
\includegraphics[width=0.45\linewidth]{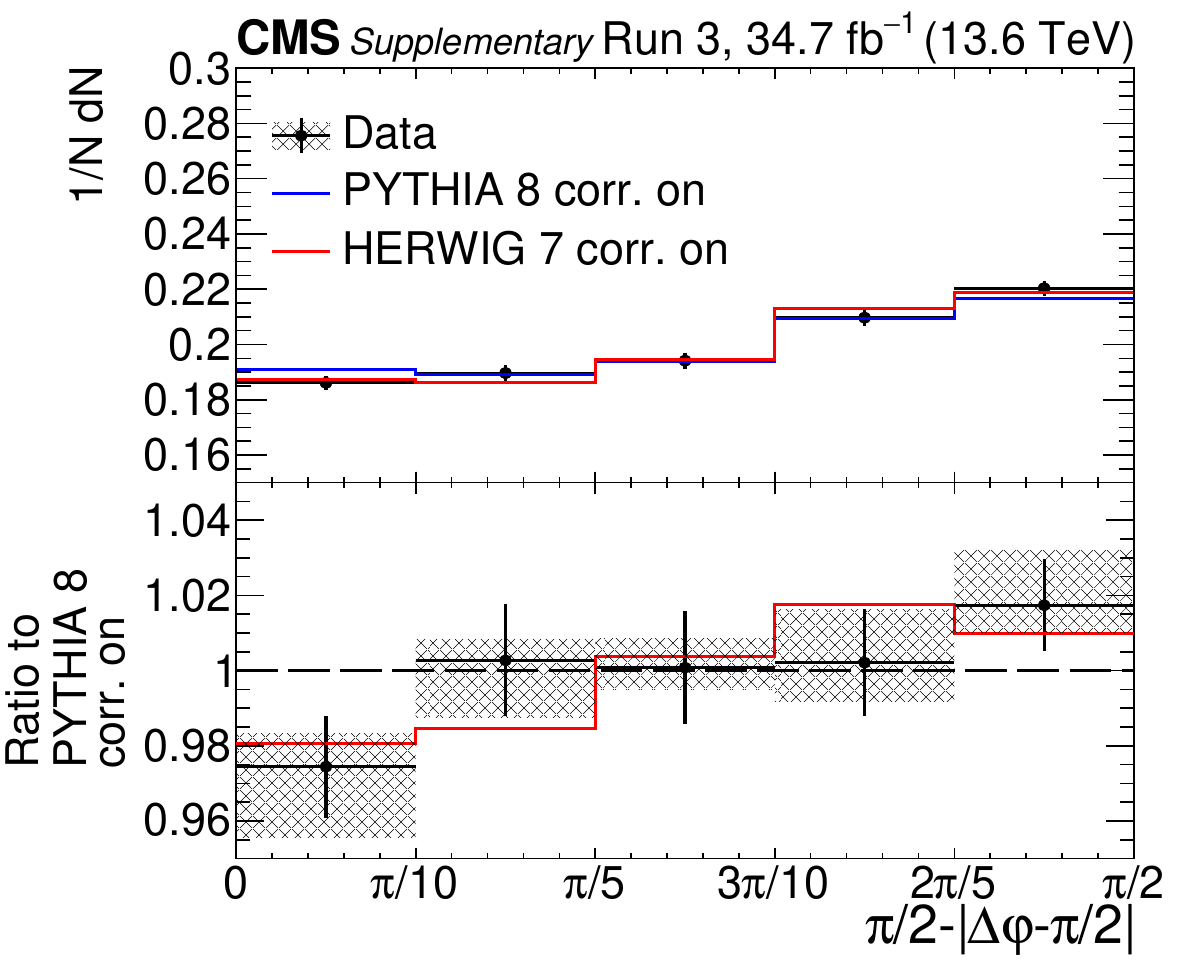}
\caption{Unfolded \dphi distribution in inclusive events without any $\sqq$ selections in data (black dots with error bars indicating
statistical uncertainties) compared with \PYTHIAviii and \HERWIGvii correlation-on predictions (histograms) passing the event selection described in the text. The lower panel shows the ratio of the data to the correlation-on \PYTHIAviii prediction. Systematic uncertainties are shown as gray hatched bands.\label{fig:unfold_inclusive}}
\end{figure*}

}
\cleardoublepage \section{The CMS Collaboration \label{app:collab}}\begin{sloppypar}\hyphenpenalty=5000\widowpenalty=500\clubpenalty=5000\cmsinstitute{Yerevan Physics Institute, Yerevan, Armenia}
{\tolerance=6000
A.~Hayrapetyan, V.~Makarenko\cmsorcid{0000-0002-8406-8605}, A.~Tumasyan\cmsAuthorMark{1}\cmsorcid{0009-0000-0684-6742}
\par}
\cmsinstitute{Institut f\"{u}r Hochenergiephysik, Vienna, Austria}
{\tolerance=6000
W.~Adam\cmsorcid{0000-0001-9099-4341}, L.~Benato\cmsorcid{0000-0001-5135-7489}, T.~Bergauer\cmsorcid{0000-0002-5786-0293}, M.~Dragicevic\cmsorcid{0000-0003-1967-6783}, P.S.~Hussain\cmsorcid{0000-0002-4825-5278}, M.~Jeitler\cmsAuthorMark{2}\cmsorcid{0000-0002-5141-9560}, N.~Krammer\cmsorcid{0000-0002-0548-0985}, A.~Li\cmsorcid{0000-0002-4547-116X}, D.~Liko\cmsorcid{0000-0002-3380-473X}, M.~Matthewman, J.~Schieck\cmsAuthorMark{2}\cmsorcid{0000-0002-1058-8093}, R.~Sch\"{o}fbeck\cmsAuthorMark{2}\cmsorcid{0000-0002-2332-8784}, M.~Shooshtari\cmsorcid{0009-0004-8882-4887}, M.~Sonawane\cmsorcid{0000-0003-0510-7010}, W.~Waltenberger\cmsorcid{0000-0002-6215-7228}, C.-E.~Wulz\cmsAuthorMark{2}\cmsorcid{0000-0001-9226-5812}
% ...
\par}
\vskip\cmsinstskip
\dag:~Deceased\\
$^{1}$Also at Yerevan State University, Yerevan, Armenia\\
$^{2}$Also at TU Wien, Vienna, Austria\\
% ...
\end{sloppypar}
\end{document}